\documentclass[modern]{aastex62}
\usepackage{amsmath}
\usepackage{comment}
\usepackage{xcolor}

\newcommand{\pdf}{\mathrm{Pr}}

\shorttitle{THE BAYESIAN SIMULATION ARGUMENT}
\shortauthors{Kipping}

\begin{document}

\title{A BAYESIAN APPROACH TO THE SIMULATION ARGUMENT}

\correspondingauthor{David Kipping}
\email{dkipping@astro.columbia.edu}

\author[0000-0002-4365-7366]{David Kipping}
\affil{Department of Astronomy,
Columbia University,
550 W 120th Street,
New York, NY 10027, USA}
\affil{Center for Computational Astophysics,
Flatiron Institute,
162 5th Av.,
New York, NY 10010, USA}



\begin{abstract}
The Simulation Argument posed by \citet{bostrom:2003} suggests that we may be
living inside a sophisticated computer simulation. If posthuman civilizations
eventually have both the capability and desire to generate such Bostrom-like
simulations, then the number of simulated realities would greatly exceed the
one base reality, ostensibly indicating a high probability that we do not live
in said base reality. In this work, it is argued that since the hypothesis
that such simulations are technically possible remains unproven, then statistical
calculations need to consider not just the number of state spaces, but the
intrinsic model uncertainty. This is achievable through a Bayesian treatment
of the problem, which is presented here. Using Bayesian model averaging, it is
shown that the probability that we are sims is in fact less than 50\%, tending
towards that value in the limit of an infinite number of simulations. This result
is broadly indifferent as to whether one conditions upon the fact that humanity has not
yet birthed such simulations, or ignore it. As argued elsewhere, it is found
that if humanity does start producing such simulations, then this would
radically shift the odds and make it very probable that we are in fact sims. 
%
\end{abstract}


\section*{ }

\section*{ }

\section*{ }

\section*{ }

\section*{ }

\section*{ }

\section*{ }

\section*{ }

\section{Introduction}

Do we live inside a computer simulation? Skepticism about our perceptions of
reality have existed for centuries, such as the ``Buttlerfly Dream'' in 
\textit{Zhuangzi} or Plato's Cave. But the development of ever more 
sophisticated computers in the modern era has led to a resurgence of interest in
the possibility that what we perceive as reality may in fact be an illusion,
simulated in some inaccessible reality above us. \citet{bostrom:2003}
formalized this possibility in his simulation argument, suggesting that
one of three distinct propositions must be true.

\begin{enumerate}
\item ``The fraction of human-level civilizations that reach a posthuman stage (that is, one capable of running high-fidelity ancestor simulations) is very close to zero'', or
\item ``The fraction of posthuman civilizations that are interested in running simulations of their evolutionary history, or variations thereof, is very close to zero'', or
\item ``The fraction of all people with our kind of experiences that are living in a simulation is very close to one''.
\end{enumerate}

The simulation argument has grown in public attention in recent years, in part
due to well-known figures such as Elon Musk expressing support for the idea,
with statements such as ``there’s a billion to one chance we’re living in
base reality'' \citep{guardianmusk}. Perhaps as a consequence of this, the
media has often described the idea as not just a possibility but in fact a
high probability (e.g. \citet{media1,media2}), which equates to the position
of \citet{bostrom:2003} if one rejects propositions 1 and 2. However, this
conditional remains unproven, and thus propositions 1 and 2 remain viable
and consistent with our knowledge and experience.

The simulation argument is not without counter-argument. One approach to
countering the idea is to ask whether a Universe-level simulation is
even possible given our understanding of the laws of nature, in other words
advocating for proposition 1 \citep{beane:2014,ringel:2017,mitchell:2020}.
For example, \citet{ringel:2017} argue that simulating quantum systems is
beyond the scope of physical plausibility. This physical argument quickly
runs into a more metaphysical obstacle though if we concede that the
possibility that our observations and understanding of physics may in fact
be simulated. In such a case, our knowledge of physics is wholly local to the
simulation and may have no real bearing on the constraints that affect a parent
reality, whose rules and limitations may be entirely different. Moreover, such
a detailed quantum-level simulation may not even be necessary to convincingly
emulate reality. Reality could be rendered in real-time locally to
deceive\footnote{Although deception may not even necessary, since the sims
have no experience/knowledge of base reality, they cannot even judge plainly
unnatural phenomenon as unphysical.} the inhabitants, rather than attempting to
generate the entire system at once. Indeed, quantum systems need not truly be
emulated, only our perceived observations of those systems, which may be far
easier.

On this basis, it is suggested that physical arguments cannot convincingly 
reject the simulation argument. Nevertheless, they clearly highlight how it
cannot be trivially assumed that proposition 1 is unquestionably true.
What other lines of reasoning might offer insight here?

At its core, the simulation hypothesis concerns a statistical argument, so
let us consider using statistical theory to address the question. Statements
such as Elon Musk's ``billion to one chance'' have popularized the position
of high statistical confidence associated with this argument.
That position stems from a frequentist perspective: count-up
how many simulated realities there are versus non-simulated ones and use
that ratio to make deduction. However, it is flawed because it tacitly
discounts propositions 1 and 2. It is useful to combine these two
propositions into a single hypothesis since their outcome is the same,
whether by capability or choice, we will not simulate realities in which
conscious self-aware beings one day reside. For notational convenience,
let us dub this the ``physical hypothesis'', $\mathcal{H}_P$. As the last
of three mutually exclusive propositions, proposition 3 necessarily
requires that propositions 1 and 2 are false, making $\mathcal{H}_P$ 
false. We dub this alternative hypothesis as 
$\mathcal{H}_S(=\bar{\mathcal{H}_P}$), the hypothesis that Bostrom-like
simulations are run by posthuman civilizations.

Certainly, if we reject $\mathcal{H}_P$ outright, then $\mathcal{H}_S$ would
be true by deduction and thus the probability that we are the base
civilization is small via proposition 3. But this is a clearly presumptive
approach, unless one had some unambiguous evidence that could fully
exclude $\mathcal{H}_P$. A rigorous statistical treatment should weigh the
hypotheses appropriately and assign probabilities which concede this
possibility - to acknowledge our ignorance. This is accomplished through a
Bayesian framework of probability, which is presented in what follows.

\section{A Statistical Analysis of the Simulation Argument}

As argued in the introduction, an evaluation of the probability
that we live in a simulated reality is best tackled from a Bayesian
perspective. The key to Bayesian statistics is Bayes' theorem,
which relates conditional probabilities to one another.
We note that arguments based on conditional probability theory have
been made previously, such as \citet{weatherson:2003}. In this work,
a deeper Bayesian approach is sought using the methods of model selection
and model averaging, in order to more directly evaluate a quantitive
measure of the probability that we live in a simulation.

To accomplish this, it is necessary to first establish results pertaining to
the likelihood of observing reality as we perceive it (our data) for each
hypothesis under consideration (in our case $\mathcal{H}_S$ and
$\mathcal{H}_P$).

\subsection{Dreams within Dreams}

We begin with the hypothesis $\mathcal{H}_S$, which describes proposition 3.
Let us consider that there exists a base civilization, which is responsible
for creating a suite of $\lambda$ simulated realities. This base civilization
is referred to as representing generation $g=1$ and its ``daughter'' simulations
as $g=2$.

\begin{figure*}
\begin{center}
\includegraphics[width=15.5cm,angle=0,clip=true]{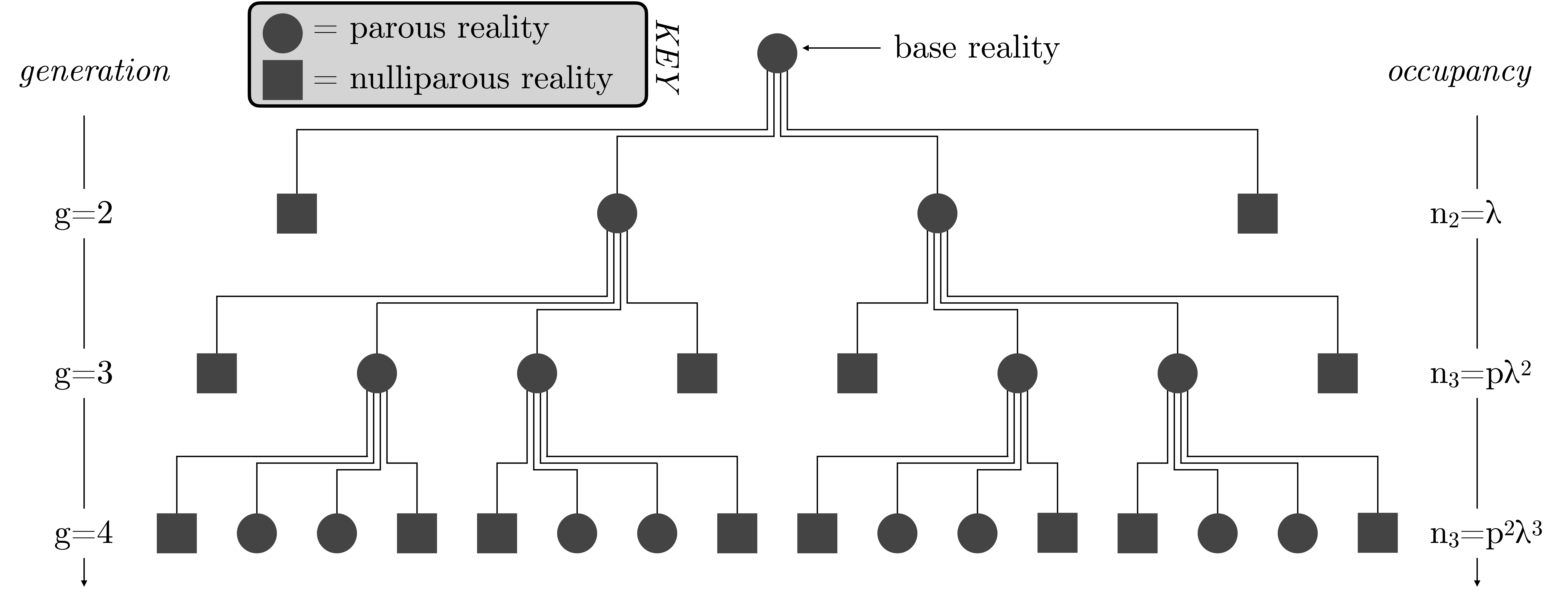}
\caption{
An illustrative depiction of a hypothetical hierarchical framework of
simulated realities embedded spawned from a base civilization.
}
\label{fig:depiction}
\end{center}
\end{figure*}

Each of these daughter simulations has a probability $p$ of itself going on to
create a suite simulations within its simulated reality, dreams within dreams.
These parous realities produce a $g=3$ generation, which can itself then create
more simulations, and so on. This creates a hierarchy of simulations, with the
base civilization - representing the only non-simulated reality - sits at the
top, as depicted in Figure~\ref{fig:depiction}.

This work treats this hierarchy as representing the ensemble of simulated realities
that will ever exist, over all times. It is suggested that it's more constructive
to pose the problem in this way, because within the simulated realities, time
itself is a construct. It doesn't really make much sense to talk about the
``passage of time'' within these realities, since they may not align to any real
world chronological definitions. They may occur almost instantaneously, played
out at some scaled version of time, or something in-between with the simulation
paused, sped-up, slowed-down at arbitrary points.

Given that our hierarchy represents the ensemble of all realities that will ever
exist, one might wonder how deep the rabbit hole goes. We suggest that there should
exist some limit to how deep the generations can indeed go, some maximum value for
$g$, denoted by $G$. This is motivated by the fact the base civilization is truly
responsible for simulating all of the daughters beneath it, and it presumably has
some finite computational limit at its disposal. By spreading this computational
power amongst $\lambda$ second-generation daughters, each of those simulations
necessarily has $1/\lambda$ less computational resources than the base
civilization devoted to creating simulations. In fact, a second-generation
civilization has even less capacity than this, since some finite fraction of this
resource is being used to generate the reality around it, besides from computers
within that reality. More generally, generation $g$ has a computational resource
of $<\lambda^{-(g-1)}$ less than that which the base civilization devotes to
simulation work.

Given that each generation has less computational power than the last, one
might suggest that this implies that $\lambda$ and $p$ should decrease as $g$
increases. This is certainly possible, but it's not strictly necessary - $p$
and $\lambda$ could remain approximately constant with respect to $g$ with the
simulations just becoming coarser in fidelity at each level, smaller in volume,
cruder in detail. This might be expected if the sims are modeled upon the
parent, with similar motivations and judgement. Despite this, even at some very
deep generation, the volume and fidelity may be perfectly sufficient to emulate
what we would recognize as reality. For this reason, it is argued here that
there isn't a good justification for invoking a variable model for $\lambda$
and $p$, which would only serve to add more complexity than warranted given our
state of knowledge of the system.

Nevertheless, the finite computational power will impose some limit on $g$,
denoted by $G$. Accordingly, from the perspective of the last generation, it is
technically impossible to ever build a computer capable of simulating
any kind reality where conscious beings reside. Their simulations would instead
be limited to more simplified programs, surely impressive still, but not
sufficient to create beings of the same conscious, self-aware experience of
reality that we enjoy.

Before proceeding, we briefly mention that although what follows is a
statistical calculation, the model is itself deterministic rather than
probabilistic. For example each parent simulation spawns the same number of
simulations. More realistically, we should expect the number to be a random
variate drawn from some underlying distribution. Whilst it may be interesting
to calculate that more mathematically challenging case, we argue here that it's
somewhat unnecessary as even this simple argument will be sufficient given the
extremely limited knowledge about the details of the simulated realities, and
again we favour invoking as simple a model as tenable given our great lack of
knowledge about simulated realities.

\subsection{Simulation Counting}

It is straight-forward to work through the first few generations and evaluate
the number of simulations in play. The second generation ($g=2$) will contain
$n_2 = \lambda$ simulations, of which $p \lambda$ will themselves be
parous. If each of these $p \lambda$ simulations yield $\lambda$ simulations
themselves, then one arrives at $n_3 = p \lambda^2$ simulations in the third
generation. Following this, one may show that in the $g^{\mathrm{th}}$
generation there are $n_g = p^{g-2} \lambda^{g-1}$ simulations.

We can now calculate the total number of simulated realities using the formula

\begin{align}
N_{\mathrm{sim}} &= \sum_{g=2}^G p^{g-2} \lambda^{g-1},\nonumber\\
\qquad&= \frac{p \lambda - (p \lambda)^G}{p-p^2\lambda},
\end{align}

where $G$ is the total number of generations. Before proceeding, it is useful
to calculate a couple of useful results using this formula. First, within the
scenario described, the vast majority of realities are of course simulated.
This really forms the basis of the oft-quoted statement that we most likely
live in a simulation. Let us write that the probability that we live in a
simulation, given a conditional denoted by CES, and also given that hypothesis
$\mathcal{H}_S$ holds, is

\begin{align}
\pdf(\mathrm{simulated}|\mathrm{CES},\mathcal{H}_S) &= \frac{N_{\mathrm{sim}}}{N_{\mathrm{sim}}+1}.
\label{eqn:frequentist}
\end{align}

What is this CES conditional? In Bayesian statistics, our inferences
are always conditioned upon some data/experience, otherwise one is simply
left with the prior beliefs. Since the simulation argument is one of
skepticism, it plausibly opens a slippery slope where essentially any
conditional information could be treated skeptically. For example, if
our data is that X humans have lived thus far prior in the pre-simulation
era, that could be challenged on the basis that our memories and records
of how many humans have lived is also simulated, and indeed our existence
could be as nascent as a few minutes ago \citep{russell:1921}. In the face
of such skepticism, the only conditional upon which we can affirm any
confidence is characterized by Ren\'e Descartes' famous ``cogito, ergo sum'' 
- CES. In this work, the conditional here really just describes the fact that
we are self-aware thinking beings that live in some kind of reality, whether
it be real or not.

We note that the number of simulations grows exponentially
with each generation, such that the last generation, $g=G$, contains a
substantial fraction of the total number, given by

\begin{align}
\pdf(g=G|\mathrm{CES},\mathcal{H}_S) &= \frac{p^{G-2} \lambda^{G-1}}{N_{\mathrm{sim}}+1},\nonumber\\
\qquad &= \Bigg( 1 - \frac{1}{p \lambda}\Bigg) \Bigg( \frac{1}{1-(p \lambda)^{1-G}} \Bigg).
\end{align}

In the limit of large $G$, or really that $(p \lambda)^G \gg 1$, this becomes

\begin{align}
\lim_{(p\lambda)^G \gg 1}\pdf(g=G|\mathcal{H}_S) &= \Bigg( 1 - \frac{1}{p \lambda}\Bigg).
\label{eqn:lowestlevel}
\end{align}

Thus, for all $p \lambda \gg 1$, most realities reside in the lowest level of the hierarchy.

\subsection{Counting Nulliparous Simulations}

Let us now consider that we have access to an additional piece of information:
we do not live in a reality that has spawned simulated realities, i.e. we are in a
nulliparous reality. This new conditional information supplants ``cogito, ergo
sum'' when available, since it implicitly includes that statement.

It is noted that \citet{poundstone:2019} qualitatively describes an ostensibly
similar conditional by distinguishing between people/sims that live before and
after the invention of simulation technology. Critically though,
\citet{poundstone:2019} consider the problem in terms of the number of
individuals that live before and after this time, which introduces analogous
issues of reference class choices and self-sampling measures
\citep{richmond:2016} that have often been central in debates concerning the
related Doomsday argument \citep{carter:1983,gott:1993,
korb:1998,bostrom:1999,simpson:2016,lampton:2020}. It also depends on the
assumption that civilizations capable of running simulations will persist
far beyond the dawn of simulation technology, so much so that most ancestor
simulations are of epochs after it's invention. No such assumptions will be
made in what follows, in an effort to cast the problem in a more agnostic
light.

As before, we continue to work under the conditional hypothesis of
$\mathcal{H}_S$. Under this hypothesis, the base civilization must be parous,
for if it were nulliparous then there would no simulated realities and
propositions 1 or 2 would be in effect, which are mutually exclusive with
proposition 3 (which in turn defines $\mathcal{H}_P$). Since the base
civilization is parous, then the conditional information that our existence is
nulliparous would immediately establish that we are not the base civilization
(under the conditional of hypothesis $\mathcal{H}_S$).

One can distinguish between two forms of nulliparity. The first is simply that
we live at the bottom of the hierarchy, the sewer of reality. After all,
Equation~(\ref{eqn:lowestlevel}) establishes that these base simulations make
up the majority of all realities\footnote{This can be thought of as an example
of applying Gott's Copernican Principle \citep{gott:1993}, if most realities
are $X$, then we most likely live in an $X$-type reality.}. In such a case,
the computational power available to the sentient beings within those realities
would simply be insufficient to feasibly ever generate daughter simulations
capable of sentient thought themselves. Here, then, proposition 2 is in effect.

As pointed out by Sean Carroll, this poses somewhat of a contradiction
\citep{carroll:2016}. If proposition 3 is true then, then it is possible to
simulate realities, and via Equation~(\ref{eqn:frequentist}) we most likely
live in a simulation, and yet more via Equation~(\ref{eqn:lowestlevel}) it is
most likely a lowest level simulation, who are incapable of simulating reality.
The conclusion that we most likely live in a reality incapable of simulating
reality, yet have assumed simulating reality is possible, forms Carroll's
contradiction. We suggest here that this contradiction can be somewhat
dissolved by considering that the lowest level may indeed not be capable of
generating their own reality simulations, but are plausibly capable of still
making very detailed simulations that fall short of generating sentience.
Accordingly, they would still suggest that it might be at least possible to
simulate realities and arrive at Bostom's trilemma all the same.

The less likely possibility is that we live in some higher level, but one of
the realities that hasn't produced a daughter - which could be because of
either proposition 1 or 2.

The total number of simulated realities between level $g=2$ and $g=G-1$
multiplied by $(1-p)$ will yield the number of nulliparous simulations which do
not reside in the lowest level, given by

\begin{align}
(1-p) \frac{p \lambda - (p \lambda)^{G-1}}{p-p^2\lambda}.
\end{align}

Adding this to the full membership of the lowest generation of the hierarchy
yields the total number of nulliparous simulated realities:

\begin{align}
N_{\mathrm{nulliparous}} &= (1-p) \frac{p \lambda - (p \lambda)^{G-1}}{p-p^2\lambda} + p^{G-2} \lambda^{G-1}.
\end{align}

Accordingly, nulliparous simulations represent the following fraction of all
realities


\begin{align}
\pdf(\mathrm{nulliparous}|\mathcal{H}_S) &= \frac{N_{\mathrm{nulliparous}}}{N_{\mathrm{sim}}+1},\nonumber\\
\qquad&= \frac{\lambda-1}{\lambda} + \frac{p (p\lambda-1)}{\lambda [(p\lambda)^G - p (1+(1-p)\lambda)]},
\end{align}

where one can see that in the limit of large $G$ this just becomes

\begin{align}
\lim_{G \to \infty} \pdf(\mathrm{nulliparous}|\mathcal{H}_S) &= \frac{\lambda-1}{\lambda}.
\label{eqn:nullS}
\end{align}

\subsection{The Physical Hypothesis}

We have now derived the necessary results to evaluate hypothesis
$\mathcal{H}_S$, but when evaluating models in Bayesian statistics it is always
necessary to compare it to some alternative(s) - in our case $\mathcal{H}_P$.
In this hypothesis, $\mathcal{H}_P$, the probability that we are nulliparous is
unity by construction of the hypothesis' definition:

\begin{align}
\pdf(\mathrm{nulliparous}|\mathcal{H}_P) &= 1.
\label{eqn:nullP}
\end{align}

Similarly, it's trivial to also write that

\begin{align}
\pdf(\mathrm{CES}|\mathcal{H}_P) &= 1.
\end{align}

\subsection{Bayes Factors Conditioned Upon Nulliparity}

Let us finally turn to evaluating a Bayes factor, the metric of Bayesian
model comparison, between the two models. In what follows, we use the
nulliparous observation as a piece of information to condition our inference
upon, but will relax this after to explore its impact.

In general, we can write the odds ratio between hypotheses $\mathcal{H}_S$ and
$\mathcal{H}_P$, conditioned upon some data $\mathcal{D}$, as

\begin{align}
O_{S:P} &= \frac{\pdf(\mathcal{H}_S|\mathcal{D})}{\pdf(\mathcal{H}_P|\mathcal{D})},\nonumber\\
\qquad&= \underbrace{\frac{\pdf(\mathcal{D}|\mathcal{H}_S)}{\pdf(\mathcal{D}|\mathcal{H}_P)}}_{\mathrm{Bayes\,\,factor}} \frac{\pdf(\mathcal{H}_S)}{\pdf(\mathcal{H}_P)}.
\end{align}

The prior ratio is generally set to unity for models with no \textit{a-priori}
preference between them, such that the odds ratio equals the Bayes factor.
This is sometimes dubbed the ``Principle of Indifference'', argued for by
Pierre-Simon Laplace, and can be thought of as a vague prior.
In our case, the ``data'' we leverage is that we are nulliparous when it
comes to simulating realities. We may thus write out the Bayes factor
in the above using Equations~(\ref{eqn:nullS}) \& (\ref{eqn:nullP})
to give

\begin{align}
\frac{\pdf(\mathrm{nulliparous}|\mathcal{H}_S)}{\pdf(\mathrm{nulliparous}|\mathcal{H}_P)} &= \frac{\lambda-1}{\lambda}.
\label{eqn:bayesfactor}
\end{align}

Since $\lambda$ can be an arbitrarily large number, then this implies
the Bayes factor is close to unity. In other words, it's approximately just
as likely that hypothesis $\mathcal{H}_S$ is true as the physical hypothesis,
given the fact we live in a nulliparous reality. However, since $\lambda$
is always a finite number, then in fact the Bayes factor is $< 1$, which means
that there is a slight preference for the physical hypothesis.

\subsection{Understanding the transition from near-certainty to ambiguity}

When we condition our inference on the fact that we are a nulliparous reality,
and employ a simple but instructive model describing a hierarchical simulated
reality, we find that the Bayes factor is close to unity. In other words, there is
no statistical preference for the simulation hypothesis over the null hypothesis
of a physical reality.

So what changed in the statistical reasoning presented here versus the more commonly
quoted conclusion that we are statistically very likely to live in a simulated
reality? After all, this is a rather dramatic turn around in conclusion given that
the simulation hypothesis has conventionally been framed as a statistical argument
and that is the line of reasoning used in this work.

There are two modifications to our thinking here than are not usually described
in arguments regarding the simulation hypothesis. The first is that we have
included this extra conditional information of nulliparity. The second is that
we have used Bayesian statistics. So we briefly consider these in turn to
evaluate where the argument changed.

\subsubsection{Neglecting our nulliparity}

Our existence as a nulliparous reality has been used as the data upon which our
Bayesian inference is conditioned, but let's now ignore that data and repeat
the Bayes factor calculation without it, to see how the conclusions change. In
the absence of this information, what other information are we going to
condition our inference upon? The only real ``data'' left is CES.

Before we can write down the revised Bayes factor, we first need to ask, what
is the probability of finding ourselves in a reality under the simulation
hypothesis i.e. $\pdf(\mathrm{CES}|\mathcal{H}_S)$? Of course the answer is
one, all simulations in the hierarchy are realities from the perspective of the
inhabitants. This conditional was implicitly present in the previous inference
too but now we explicitly write it out since there is an absence of any other
information. Similarly, for the physical hypothesis, we have
$\pdf(\mathrm{CES}|\mathcal{H}_P) = 1$.

This means that the Bayes factor is straight-forwardly:

\begin{align}
\frac{\pdf(\mathrm{CES}|\mathcal{H}_S)}{\pdf(\mathrm{CES}|\mathcal{H}_P)} &= 1.
\label{eqn:bayesfactoralt}
\end{align}

If we compare this to Equation~(\ref{eqn:bayesfactor}), it's almost
identical - there too we obtained nearly even odds between the two
hypotheses. Therefore, this reveals that our modification of including the
nulliparous information is not responsible for the revised conclusion
of ${\sim}$50:50 odds for the simulation hypothesis. By deduction, that
indicates that it is the \textit{Bayesian} treatment of probabilities
that must be responsible.

\subsubsection{Negating Bayesian model comparison}

To demonstrate this, let's see if we can recover the often claimed
conclusion that statistically we are more likely to live a simulated
reality by sheer numbers. This is straight-forward to see when
one operates under the tacit assumption that the $\mathcal{H}_S$
is true. If we assert that is true, then the vast majority of realities
are indeed simulated. But the fallacy of this argument is that
we have already assumed it's correct, whereas the Bayes factors derived
earlier compare the hypothesis that it is/it is not true. This is the
key difference driving the radically different conclusions.

To show this, let's now treat $\mathcal{H}_S$ as a fixed conditional.
We can no longer ask if the simulation hypothesis, as defined by
$\mathcal{H}_S$ is true, since it is asserted as so. Instead, we ask
what is the probability that $g=1$ (we are the first generation in
the hierarchy) given that we exist?

\begin{align}
\pdf(g=1|\mathrm{CES},\mathcal{H}_S) &= \frac{1}{N_{\mathrm{sim}}+1}
\label{eqn:g1}
\end{align}

Since the number of simulations can be very large, this recovers the
conventional conclusion that we are almost certainly living in a simulation.
This is really just a frequentist argument and is clearly conditional upon
the simulation hypothesis itself already being true.

We can also modify the above to include the conditional that we exist in a
nulliparous reality, as before. Given that only simulated realities exist in
such a state in the hypothesis $\mathcal{H}_S$, then this simply yields
$\pdf(g=1|\mathrm{nulliparous},\mathcal{H}_S)=0$.

\subsection{Bayesian Model Averaging with the ``Cogito, Ergo Sum'' Conditional}

Thus far we have evaluated
i] the probability that we are the base reality when proposition 3/hypothesis
$\mathcal{H}_S$ is true (a very small number);
ii] the probability that we are in the base reality when $\mathcal{H}_P$
is true (which is trivially 1); and
iii] the Bayes factor between hypotheses $\mathcal{H}_S$ and $\mathcal{H}_P$,

The latter doesn't quite provide a direct answer to the question as to whether
we live inside a simulation though, since one of the realities embedded within
$\mathcal{H}_S$ is real - namely the base reality. However, ideally we would
combine these three results to evaluate the probability that $g=1$ marginalized
over our uncertainty about which hypothesis is correct.

This can actually be formalized through the use of Bayesian model averaging.
Let's say we wish to evaluate the probability that $g=1$ (i.e. we are
the first generation) as we did in Equation~(\ref{eqn:g1}), but now we wish to
relax the conditional assumption made earlier that $\mathcal{H}_S$ is
assumed to be true. Instead, we can calculate the probability that $g=1$
for both hypotheses weighted by their model evidence in their favour, known
as Bayesian model averaging. By doing so, we incorporate our ignorance about
which model is correct. This essentially looks like a discrete marginalization
over hypothesis-space:

\begin{align}
\pdf(g=1|\mathrm{CES}) =&
\pdf(g=1|\mathrm{CES},\mathcal{H}_S) \pdf(\mathcal{H}_S|\mathrm{CES}) + \nonumber\\
\qquad& \pdf(g=1|\mathrm{CES},\mathcal{H}_P) \pdf(\mathcal{H}_P|\mathrm{CES}).
\end{align}

We may now use results from earlier, including Equation~(\ref{eqn:g1}),
to re-write this as

\begin{align}
\pdf(g=1|\mathrm{CES}) =&
\frac{\pdf(\mathcal{H}_S|\mathrm{CES})}{N_{\mathrm{sim}}+1} + \pdf(\mathcal{H_P}|\mathrm{CES}).
\label{eqn:temp1}
\end{align}

From Equation~(\ref{eqn:bayesfactoralt}), we have

\begin{align}
\frac{ \pdf(\mathrm{CES}|\mathcal{H}_S) }{ \pdf(\mathrm{CES}|\mathcal{H}_P) } = 1,
\end{align}

and so using Bayes' theorem this becomes

\begin{align}
\frac{ \pdf(\mathcal{H}_S|\mathrm{CES}) }{ \pdf(\mathcal{H}_P|\mathrm{CES}) } \frac{ \pdf(\mathcal{H}_S) }{ \pdf(\mathcal{H}_P) } = 1.
\end{align}

To make progress, we must assign the prior hypothesis probabilities, which is typically
simply set to unity as an uninformative choice, giving

\begin{align}
\pdf(\mathcal{H}_S|\mathrm{CES}) - \pdf(\mathcal{H}_P|\mathrm{CES}) = 0.
\end{align}

We also exploit the fact that the sum of all probabilities must equal one, such that

\begin{align}
\pdf(\mathcal{H}_S|\mathrm{CES}) + \pdf(\mathcal{H}_P|\mathrm{CES}) = 1.
\end{align}

Simultaneously solving the last two equations gives

\begin{align}
\pdf(\mathcal{H}_S|\mathrm{CES}) = \tfrac{1}{2},\nonumber\\
\pdf(\mathcal{H}_P|\mathrm{CES}) = \tfrac{1}{2}.
\end{align}

And now finally plugging this back into Equation~(\ref{eqn:temp1}) gives

\begin{align}
\pdf(g=1|\mathrm{CES}) =&
\frac{1}{2} + \frac{1}{2 (N_{\mathrm{sim}}+1)}.
\end{align}

From the above, we have that the probability that we live in the base reality,
$g=1$, is one-half plus some additional term which depends on the number of
simulations in the simulation hypothesis, $N_{\mathrm{sim}}$. Since
$N_{\mathrm{sim}}\geq0$, then $\pdf(g=1|\mathrm{CES})\leq1$ as expected. It
also means that $\pdf(g=1|\mathrm{CES})>\tfrac{1}{2}$ for all
$N_{\mathrm{sim}}$, asymptotically tending towards one half in the limit of
large $N_{\mathrm{sim}}$. On this basis, it is in fact more likely that we live
in the $g=1$ base reality than a simulation, although it may be only very
slightly more preferable depending on one's assumptions for $N_{\mathrm{sim}}$.

\subsection{Bayesian Model Averaging with the Nulliparous Conditional}

For completion, we will now repeat the previous subsection but replace
the conditional ``cogito ergo sum'' with the nulliparous case.

\begin{align}
\pdf(g=1|\mathrm{nulliparous}) =&
\pdf(g=1|\mathrm{nulliparous},\mathcal{H}_S) \pdf(\mathcal{H}_S|\mathrm{nulliparous}) + \nonumber\\
\qquad& \pdf(g=1|\mathrm{nulliparous},\mathcal{H}_P) \pdf(\mathcal{H}_P|\mathrm{nulliparous})
\end{align}

In this case the likelihoods are binary, giving

\begin{align}
\pdf(g=1|\mathrm{nulliparous}) =&
0 \times \pdf(\mathcal{H}_S|\mathrm{nulliparous}) + 1 \times \pdf(\mathcal{H}_P|\mathrm{nulliparous}).
\end{align}

The Bayes factor between the two models, from Equation~(\ref{eqn:bayesfactor}),
can be written (in the limit of large $G$) as

\begin{align}
\frac{ \pdf(\mathcal{H}_S|\mathrm{nulliparous}) }{ \pdf(\mathcal{H}_P|\mathrm{nulliparous}) } \frac{ \pdf(\mathcal{H}_S) }{ \pdf(\mathcal{H}_P) } = \frac{\lambda-1}{\lambda},
\end{align}

which again we simplify by invoking even a-priori odds between the models

\begin{align}
\lambda \pdf(\mathcal{H}_S|\mathrm{nulliparous}) = (\lambda-1) \pdf(\mathcal{H}_P|\mathrm{nulliparous}).
\end{align}

We combine this with the fact that sum of the probabilities equals one, as before,
to write that

\begin{align}
\pdf(g=1|\mathrm{nulliparous}) =& \pdf(\mathcal{H}_P|\mathrm{nulliparous}),\nonumber\\
\qquad=& \frac{1}{2-\lambda^{-1}}.
\end{align}

Note that in the simulation hypothesis, within which $\lambda$ finds its definition,
that the $\lambda$ term is $\geq1$. Examination of our formula indeed reveals, as
before that, $\tfrac{1}{2}<\pdf(g=1|\mathrm{nulliparous})\leq1$, with the formula
asymptotically tending towards a half (but always remaining greater than it) for
large $\lambda$. As also with before then, the conclusion is that we are more likely
to be living in the $g=1$ reality, although perhaps only with slight preference.

\subsection{What if we were parous?}

Although we are presently not a parous reality, it is interesting to consider
how the results derived thus far would change if tomorrow we began producing 
Bostrom-like simulations. For hypothesis $\mathcal{H}_P$, it is simple to write
that 

\begin{align}
\pdf(\mathrm{parous}|\mathcal{H}_P) = 0.
\label{eqn:parousHp}
\end{align}

Through Bayesian model averaging, we have that the probability of
$g=1$, with the condition of being parous, is given by

\begin{align}
\pdf(g=1|\mathrm{parous}) =&
\pdf(g=1|\mathrm{parous},\mathcal{H}_S) \pdf(\mathcal{H}_S|\mathrm{parous}) + \nonumber\\
\qquad& \pdf(g=1|\mathrm{parous},\mathcal{H}_P) \pdf(\mathcal{H}_P|\mathrm{parous}).
\end{align}

The second row goes to zero since $\pdf(\mathcal{H}_P|\mathrm{parous}) \propto
\pdf(\mathrm{parous}|\mathcal{H}_P) = 0$ via Equation~(\ref{eqn:parousHp}).

\begin{align}
\pdf(g=1|\mathrm{parous}) =&
\pdf(g=1|\mathrm{parous},\mathcal{H}_S) \pdf(\mathcal{H}_S|\mathrm{parous})
\end{align}

Since $\pdf(\mathcal{H}_P|\mathrm{parous})=0$, then
$\pdf(\mathcal{H}_S|\mathrm{parous})=1$ by requirement that the probabilities
sum to one, and so

\begin{align}
\pdf(g=1|\mathrm{parous}) =& \pdf(g=1|\mathrm{parous},\mathcal{H}_S).
\end{align}

Under hypothesis $\mathcal{H}_S$, we have already calculated that a fraction
$(\lambda-1)/\lambda$ are nulliparous via Equation~(\ref{eqn:nullS}). Accordingly,
one minus this are parous, which equals $1/\lambda$. The total number of parous
realities is therefore $N_{\mathrm{sim}}/\lambda$. Only one of these is the base
reality, and so we have

\begin{align}
\pdf(g=1|\mathrm{parous}) =& \frac{\lambda}{N_{\mathrm{sim}}}.
\end{align}

For large $G$, $N_{\mathrm{sim}}\gg\lambda$ and thus this probability approaches
zero. Accordingly, if we become a parous reality, in other we start producing
Bostrom-like simulations, the probability that we live in a simulated reality
radically shifts from just below one-half to just approaching zero.

\section{Discussion}
\label{sec:discussion}

In this work, we have divided the three propositions of \citet{bostrom:2003}
into two hypotheses: one where simulated realities are produced
($\mathcal{H}_S$), and one where they are not ($\mathcal{H}_P$). Comparing the
models with Bayesian statistical methods, it is found that that the Bayes
factor is approximately unity, with a slight preference towards
$\mathcal{H}_P$. Whilst the Bayes factor can be objectively stated without the
need to assign any priors, the odds ratio between the two models depends on the
prior model probabilities, $\pdf(\mathcal{H}_S)/\pdf(\mathcal{H}_P)$. A
standard choice is to assume all models are \textit{a-priori} as likely as each
other, but this could be challenged as being too generous to model
$\mathcal{H}_S$, on the basis that it is an intrinsically far more complex
model.

If one goes further and assigns a value to the ratio of the prior model
probabilities, then one can use Bayesian model averaging to marginalize
over the models, weighted by their posterior probabilities. If one does not
penalize the model $\mathcal{H}_S$ for its complexity and simply assigns
even \textit{a-priori} odds, then it is still found that the probability
we live in base reality - after marginalizing over the model uncertainties -
is still not the favored outcome, with a probability less than 50\%.
As the number of simulations grows very large, this probability tends towards
50\%, and thus it is argued here that the most generous probability that
can be assigned to the idea that we live inside a simulation is one half.

It is argued that the results presented are robust against the choice of
self sampling. For example, if one replaced the conditionals used here, which
describe the reality in which one finds oneself, with the number of sims
in each state/reality, this would not noticeably affect the results under the
assumption that the number of sims per reality is evenly distributed. This is
because our results asymptotically tend towards 50\% in the case of a large
number of realities, but that would be equally so if one replaced realities
with sims instead, which would display the same asymptotic behaviour.

\section*{Acknowledgements}

DMK is supported by the Alfred P. Sloan Foundation. Thank-you to reviewer \#1 for such a thoughtful report
of this work. DK would like to thank Tom Widdowson, Mark Sloan, Laura Sanborn, Douglas
Daughaday, Andrew Jones, Jason Allen, Marc Lijoi, Elena West \& Tristan
Zajonc.



\end{document}